\documentclass[reprint,aps,floatfix,10pt]{revtex4-1}
\usepackage{amsmath,amssymb}
\usepackage{graphicx}
\usepackage[scale=0.7, centering, margin=2cm, includefoot, a4paper]{geometry}

\def\uu{{\bf u}}

\def\be{\begin{equation}}
\def\ee{\end{equation}}
\def\ba{\begin{eqnarray}}
\def\ea{\end{eqnarray}}
\usepackage[outdir=./]{epstopdf}
\usepackage{color}

\begin{document}

\title{Statistics of incompressible hydrodynamic turbulence: an alternative approach}

\author{Nahuel Andr\'es}
\email{nahuel.andres@lpp.polytechnique.fr}
\affiliation{Laboratoire de Physique des Plasmas, Palaiseau 91128, France}
\author{Supratik Banerjee}
\email{sbanerjee@iitk.ac.in}
\affiliation{Department of Physics, Indian Institute of Technology Kanpur, Kalyanpur 208016, Uttar Pradesh, India}
	
\date{\today}

\begin{abstract}
Using a recent alternative form of the Kolmogorov-Monin exact relation for fully developed hydrodynamics (HD) turbulence, the incompressible energy cascade rate $\varepsilon$ is computed. Under this current theoretical framework, for three-dimensional (3D) freely decaying homogeneous turbulence, the statistical properties of the fluid velocity ($\uu$), vorticity ($\boldsymbol \omega= \boldsymbol\nabla \times \uu$) and Lamb vector {($\boldsymbol{\cal L}= \boldsymbol \omega \times \uu)$} are numerically studied. For different spatial resolutions, the numerical results show that $\varepsilon$ can be obtained directly as the simple products of two-point increments of $\uu$ and $\boldsymbol{\cal L}$, without the assumption of isotropy. Finally, the results for the largest spatial resolutions show a clear agreement with the cascade rates computed from the classical 4/3 law for isotropic homogeneous HD turbulence.  
\end{abstract}


\maketitle
	
\section{Introduction}

Turbulence is a non-linear phenomenon omnipresent in nature. However, due to extremely complex nature, its full understanding remains far to be completed. For fully developed turbulence, the fluid flow contains fluctuations populating a wide range of space- and time-scales. In the so-called inertial range, sufficiently decoupled from the injection/forcing large-scales and the dissipation small-scales, the kinetic energy (or other inviscid invariants of the flow) takes part in a cascade process across the different scales. This process is characterized by a scale independent cascade rate, i.e. $\varepsilon$, which represents the universality of turbulence.

In the theory of statistically homogeneous turbulence \citep{F1995}, there are only a few number of exact results. For three-dimensional (3D), homogeneous, isotropic and incompressible HD turbulence, in the limit of infinitely large kinetic Reynolds number, one of the most important exact results is the so-called 4/5 law. This type of exact laws are crucial for obtaining an accurate and quantitative estimate of the energy dissipation rate $\varepsilon$, and hence, of the heating rate by the process of the turbulent cascade. In its anisotropic generalization, the so-called Kolmogorov-Monin relation, can be cast as \citep{MY1975},
\begin{equation}\label{original}
	-2 \varepsilon = \boldsymbol \nabla_{\boldsymbol\ell} \cdot \left\langle  |\delta \uu|^2 \delta  \uu  \right\rangle,
\end{equation}
where $\delta {\bf u} \equiv {\bf u}({\bf x}+\boldsymbol\ell)-{\bf u}({\bf x})$ is the velocity increment, ${\bf x}$ is a reference point and $\boldsymbol\ell$ is the separation vector. It is worth mentioning that Eq.~\eqref{original} expresses the energy cascade rate $\varepsilon$ purely in terms of the two-point third-order structure functions \citep[see, e.g.][]{vkh1938,K1941a,MY1975,F1995}. In practice, also, one has to integrate the above equation in order to calculate $\varepsilon$ from numerical or observational data. When isotropy is assumed, the integrated form of Eq.~\eqref{original} predicts a linear scaling between the third-order velocity structure function and the seperation length scale $\ell$  \citep[see reference therein, ][]{A2016b}. As a consequence, this scaling law, and in general all scaling laws, put strong boundaries to the theories of turbulence. Similar analytical relations have also been derived using different models of incompressible (and compressible) plasma turbulence, with and without the assumption of isotropy \citep{P1998a,P1998b,P2003,Ga2008,B2013,A2016b,A2016c}. However, for an anisotropic or compressible flow, the computation of $\varepsilon$ becomes much more difficult because of the absence of spherical symmetry \citep[see,][]{A2018b} or the presence of source/sink terms in the exact law \citep{Ga2011,B2013,B2014,A2017b,A2018,BF2018}. 

Recently, inspired by the Lamb formulation \citep{L1877}, a number of non-conventional exact laws have been derived for fully developed turbulence \citep{B2017}. Using two-point statistics, \citet{B2017} have found that the {energy cascade rate} can be expressed simply in terms of second-order mixed structure functions. In particular, in this simpler algebraic form the authors have found that the Lamb vector, i.e. ${\cal L} \equiv \boldsymbol\omega\times\uu$, plays a key role in the HD turbulent process. Moreover, unlike Eq.~\eqref{original}, the alternative exact relation gives directly $\varepsilon$ without going through an integration. Hence, the current form is equally valid for a turbulent flow with and without the assumption of isotropy. The main objective of the present paper is to calculate $\varepsilon$ using the recently derived alternative exact law for incompressible HD turbulence. For our study, we use numerical data obtained from 3D direct numerical simulations (DNSs) with spatial resolution ranging from $128^3$ to $1536^3$ grid points. In the course of this study, we also investigate the statistical behavior of the velocity, vorticity and the Lamb vector fluctuations.

The paper is organized as follows: in Sec.~\ref{equations} we describe the equations and the code used in the present work; in {Sec.~\ref{classical} and \ref{alternative} we present the classical and alternative exact laws} for fully developed HD turbulence. In particular, we present a brief analysis of the exact law, with a particular emphasis on the structure  of each term involves in the nonlinear cascade of energy; in Sec.~\ref{results} we present our main numerical results; and, finally, in Sec.~\ref{discussion} we discuss the main findings and their implications.

\section{Theory and Numerical Simulations}

\subsection{Navier-Stokes Equation $\&$ Code}\label{equations}

We solve numerically the equations for an incompressible fluid with constant mass density and without external forcing. Then, the Navier-Stokes equation reads,
\begin{equation}\label{n-s}
	\frac{\partial \uu}{\partial t} = - (\uu\cdot\boldsymbol\nabla) \uu -\boldsymbol\nabla p + \nu\nabla^2 \uu, 
\end{equation}
with the constrain $\boldsymbol\nabla\cdot\uu = 0$, $p$ is the scalar pressure (normalize to the constant unity density) and $\nu$ is the kinematic viscosity. In the present paper, our numerical results steam from the analysis of a series of DNSs of Eq.~\eqref{n-s} using a parallel pseudo-spectral code in a three-dimensional box of size $2\pi$ with periodic boundary conditions, from $N=128$ up to $N=1536$ linear grid points. The equations are evolved in time using a second order Runge-Kutta method, and the code uses the 2/3-rule for dealiasing \citep{Go2005, Mi2008, Mi2011}. As a result, the maximum wave number for each simulation is $k_{max}=N/3$, where $N$ is the number of linear grid points. We can define the viscous dissipation wave number as $k_\eta = (\langle\omega^2\rangle/\nu^2)^{1/4}$, and as a consequence the Kolmogorov scale is equal to $\eta=2\pi/k_\eta$. It is worth mentioning that all simulations presented are well resolved, i.e. the dissipation wave number $k_\eta$ is smaller than the maximum wave number $k_{max}$ at times where the statistical computations have been done. 

The initial state in our simulations consists of isotropic velocity field fluctuations with random phases, such that the total helicity is zero, and the kinetic energy initially is equal to $1/2$ and localized at the largest scales of the system (only wavenumber $k = 2$ is initially excited). There is no external forcing and our statistical analysis is made at a time when the mean dissipation rate reaches its maximum (around 5 turnover times). We also can define the Taylor and integral scale as,
\ba\label{scales}\label{lambda}
    \lambda &=& 2\pi \left(\frac{\int E(k)dk}{\int E(k)k^2dk}\right)^{1/2}, \\\label{integral}
	L &=& 2\pi \frac{\int E(k)k^{-1}dk}{\int E(k)dk},
\ea
where $E(k)$ is the kinetic energy spectrum. From the definitions \eqref{lambda} and \eqref{integral}, we can compute the corresponding Reynolds number $R_L=U_0L/\nu$ and the Taylor-based Reynolds number $R_\lambda=U_0\lambda/\nu$ (here, $U_0=\langle u^2 \rangle^{1/2}$ is the rms velocity). Table \ref{table} summarized these values for all Runs used in the present paper.

\begin{table*}[t]
	\begin{tabular}{p{1.2cm} p{1.2cm} p{1.5cm} p{1.2cm} p{1.2cm} p{1.2cm} p{1.2cm} p{1.2cm} p{1.2cm} p{1.2cm}}
	\hline\hline
	 Run & $N$ & $\nu$ & $\lambda$ & $L$ & $\langle u^2 \rangle^{1/2}$ & $\langle \omega^2 \rangle^{1/2}$ & $R_\lambda$ & $R_L$ & $k_{max}/k_\nu$ \\
	\hline
 		 I  & 128 & $3.0\times10^{3}$ & 0.99 & 2.49  & 0.78 & 5.31 & 258 & 646 & 1.02 \\
 		II  & 256 & $1.5\times10^{3}$ & 0.83 & 2.38 & 0.76 & 7.99 & 419 & 1205 & 1.17\\
       III  & 512 & $7.5\times10^{4}$ & 0.42 & 1.74 & 0.77 & 12.45 & 435 & 1789 & 1.32\\
 		IV  & 1024 & $3.0\times10^{4}$ & 0.27 & 1.60 & 0.79 & 19.28 & 725 & 4212 & 1.34\\
 		 V  & 1536 & $1.5\times10^{4}$ & 0.15 & 1.50 & 0.81 & 34.60 & 870 & 8736 & 1.03\\
	\hline\hline
    \end{tabular}
	\caption{Parameters used in Runs I to V: $N$ is the linear grid points; $\nu$ is the kinematic viscosity; $\lambda$ and $L$ are the Taylor and integral scale, respectively; $\langle u^2 \rangle^{1/2}$ and $\langle \omega^2 \rangle^{1/2}$ are the rms velocity and rms vorticity, respectively; $R_\lambda$ and $R_L$ are the Reynolds numbers based in the Taylor and integral scale, respectively and $k_{max}/k_\nu$ is the maximum to the dissipation wavenumber ratio.
    \label{table}}
\end{table*}

\begin{figure*}
\begin{center}
	\includegraphics[width=.89\textwidth]{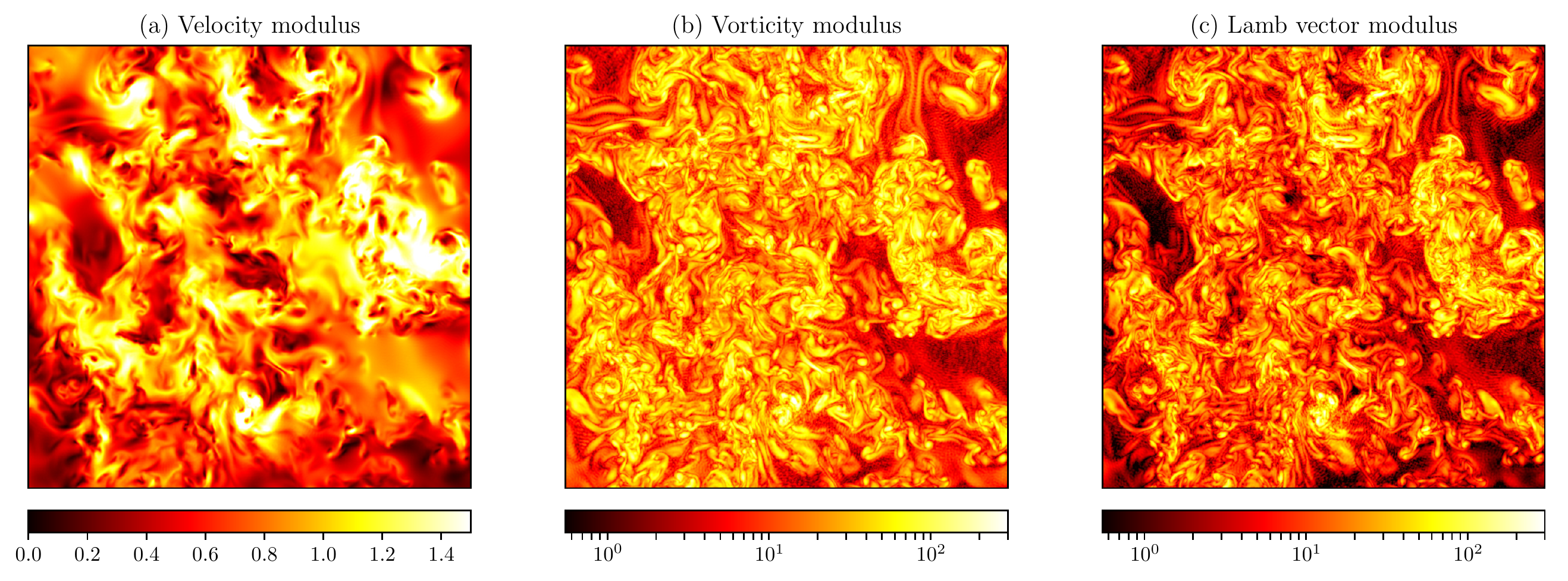}
\end{center}
\caption{Snapshot (512$\times$512) of the velocity (a), vorticity (b) and Lamb vector (c) modulus for Run V on linear and logarithm scale, respectively.}\label{fig1}
\end{figure*}

\begin{figure*}
\begin{center}
	\includegraphics[width=.3\textwidth]{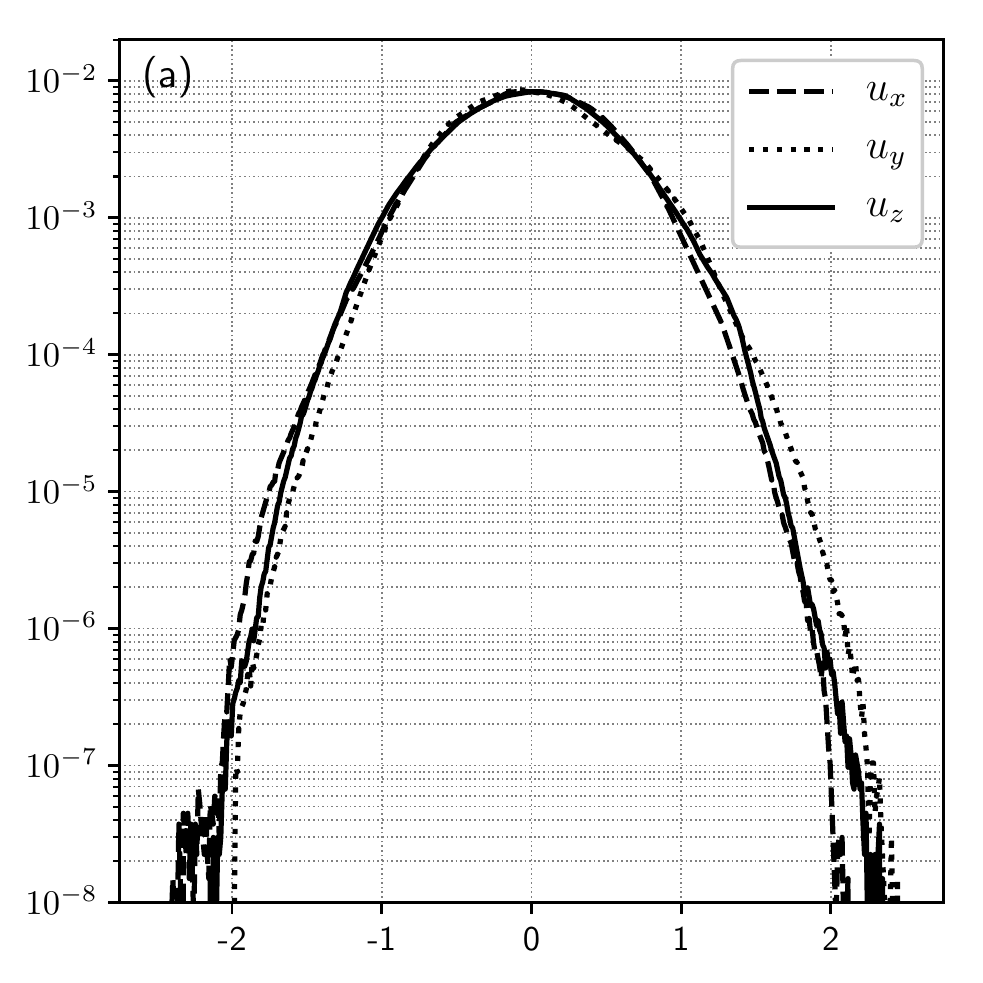}\includegraphics[width=.3\textwidth]{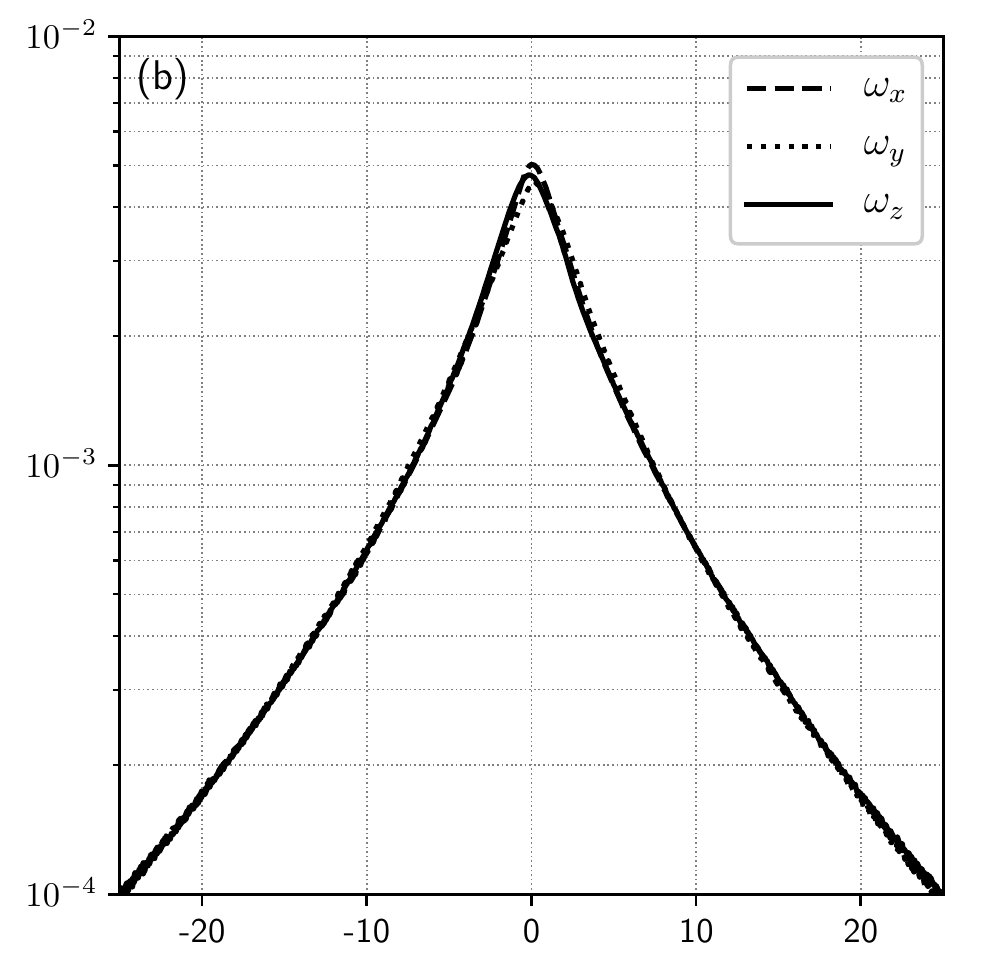}\includegraphics[width=.3\textwidth]{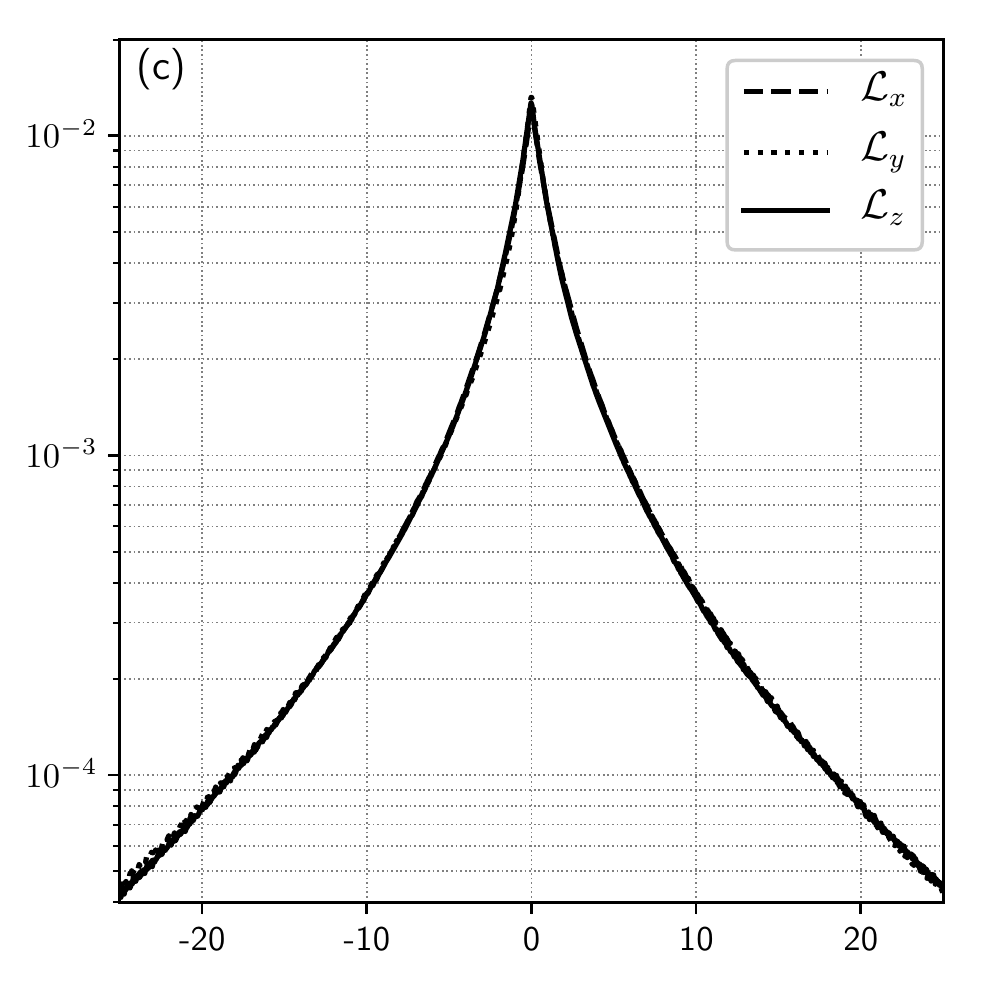}
\end{center}
\caption{For Run IV: PDFs of the velocity (a), vorticity (b) and Lamb vector (c) components.}\label{fig2}
\end{figure*}

{\subsection{Classical exact law}\label{classical}}

As we discussed in the Introduction, following the original works of Kolmogorov and Monin and Yaglom derivations \citep{K1941a,MY1975} for homogeneous and isotropic HD turbulence, {assuming statistical stationarity and a finite energy cascade rate as $\nu$ goes to zero,} we can compute the energy cascade rate as a function of the third-order velocity structure functions as,
\be\label{projection}
	-\frac{4}{3} \varepsilon \ell = \left\langle  |\delta \uu|^2 \delta  u_\ell\right\rangle = \langle F_\ell \rangle,
\ee
where $u_\ell$ is the projection of the velocity field on the increment direction $\boldsymbol{\ell}$. Eq.~\eqref{projection} is the so-called four-third law, which can be also derived from Eq.~\eqref{original} assuming isotropic turbulence. Usually, the mean flux term $\langle{F}_\ell\rangle \equiv \langle  |\delta \uu|^2 \delta u_\ell \rangle$ along $\boldsymbol{\ell}$ is identified as the flux of kinetic energy through scales. It is worth mentioning that in the new alternative derivation to compute the energy cascade rate {from \citet{B2017} (see Sec.~\ref{alternative})}, there is no projection along the increment direction $\boldsymbol{\ell}$ and the expression only depends in the two-point mixed structure functions. In particular, this would be essential when there is a privileged direction in the system, as in magnetohydrodynamics (MHD) with a magnetic guide field \citep[see, e.g.][]{Sh1983,M1996,Ga2000,W2012,Ou2013,A2017a} or in rotating HD turbulence \citep[see, e.g.][]{C2014,R2015,C2016,P2018}.

{\subsection{Alternative exact law}\label{alternative}}

{Following \citet{B2017}, here we give an schematic description of the derivation of the alternative exact relation for fully developed homogeneous and incompressible turbulence. The alternative Navier-Stokes Eq.~\eqref{n-s} can be cast as,
\begin{equation}\label{alt_n-s}
	\frac{\partial \uu}{\partial t} =  (\uu\times\boldsymbol{\omega}) -\boldsymbol\nabla p_T + \nu\nabla^2 \uu, 
\end{equation}
where the non-linear term have been written as (minus) the Lamb vector and as a part of the total pressure $p_T\equiv p+u^2/2$.} {The symmetric two-point correlators for the total energy can be defined as,
\begin{equation}
    R_E=R_E'\equiv\frac{1}{2}\langle\uu\cdot\uu'\rangle,
\end{equation}
where the prime implies variable at ${\bf x'={\bf x}+\boldsymbol{\ell}}$ point.} {Using Eq.~\eqref{alt_n-s}, the dynamical evolution equation for the energy correlator is,
\begin{equation}
    \partial_t(R_E+R_E') = -\langle\uu '\cdot \boldsymbol{\cal L}\rangle-\langle\uu \cdot\boldsymbol{\cal L}'\rangle+{\cal D}+{\cal F}, \label{algo}
\end{equation}
where we have used the constraint $\boldsymbol\nabla\cdot\uu = 0$, the relations $\boldsymbol{\cal L}\cdot\uu=0=\boldsymbol{\cal L}'\cdot\uu '$ and $\cal D, F$ represent the correlation terms related to the dissipation and forcing, respectively.} {Then, assuming the usual assumptions for fully developed turbulence (where an asymptotic stationary state is expected to be reached) \citep{B2017,A2016b,Ga2011}, we can derive an exact law valid in the inertial range. In particular, assuming an infinite kinetic Reynolds number with a statistical balance between forcing and dissipation terms and a finite energy cascade rate as we go to the zero viscosity limit, ${\bf\cal D}\sim 0$, ${\bf\cal F} \sim 2\varepsilon$ and Eq.~\eqref{algo} can be cast as,
\begin{equation}
    -2\varepsilon = -\langle\uu '\cdot\boldsymbol{\cal L}\rangle-\langle\uu \cdot \boldsymbol{\cal L}'\rangle.
\end{equation}
Finally, using statistical homogeneity, we obtain after few steps of simple Algebra, the alternative formulation of the exact law is,
\begin{equation}\label{first1}
	2 \varepsilon = - \left\langle  \delta \boldsymbol{\cal L} \cdot \delta  \uu  \right\rangle = \left\langle  \delta (\uu \times {\boldsymbol \omega}) \cdot \delta  \uu  \right\rangle,  
\end{equation}
where $\delta {\bf u} \equiv {\bf u}({\bf x}+\boldsymbol\ell)-{\bf u}({\bf x})$ is the usual increment. Eq.~\eqref{first1} gives a divergence free exact relation for homogeneous incompressible turbulence valid in the inertial range, i.e. far away from the forcing and dissipative scales. Unlike the Eq.~\eqref{original}, this new expression does not involve a third-order structure function but second-order mixed structure functions. Besides, there is no global divergence in the alternative formulation. Therefore, the estimation of the energy cascade rate can be obtained directly from the measurement of the scalar product of the Lamb vector increments with the velocity increments.}

Equation \eqref{first1} can be cast as,
\begin{equation}\label{first2}
 	2 \varepsilon = \varepsilon_x +  \varepsilon_y + \varepsilon_z, 
\end{equation}
where we have identified three specific contributions as, 
\ba\label{epsx}
\varepsilon_x &\equiv & -\left\langle \delta {\cal L}_x \delta u_x \right\rangle = \left\langle \delta (u_y \omega_z - u_z \omega_y) \delta u_x \right\rangle, \\\label{epsy}
\varepsilon_y &\equiv & -\left\langle \delta {\cal L}_y \delta u_y \right\rangle =\left\langle \delta (u_z \omega_x - u_x \omega_z) \delta u_y \right\rangle, \\ \label{epsz}
\varepsilon_z &\equiv& -\left\langle \delta {\cal L}_z \delta u_z \right\rangle = \left\langle \delta (u_x \omega_y - u_y \omega_x) \delta u_z \right\rangle.
\ea
In homogeneous and isotropic turbulence, we expect that each of these contributions be statistically the same. 

\begin{figure}
\begin{center}
	\includegraphics[width=.45\textwidth]{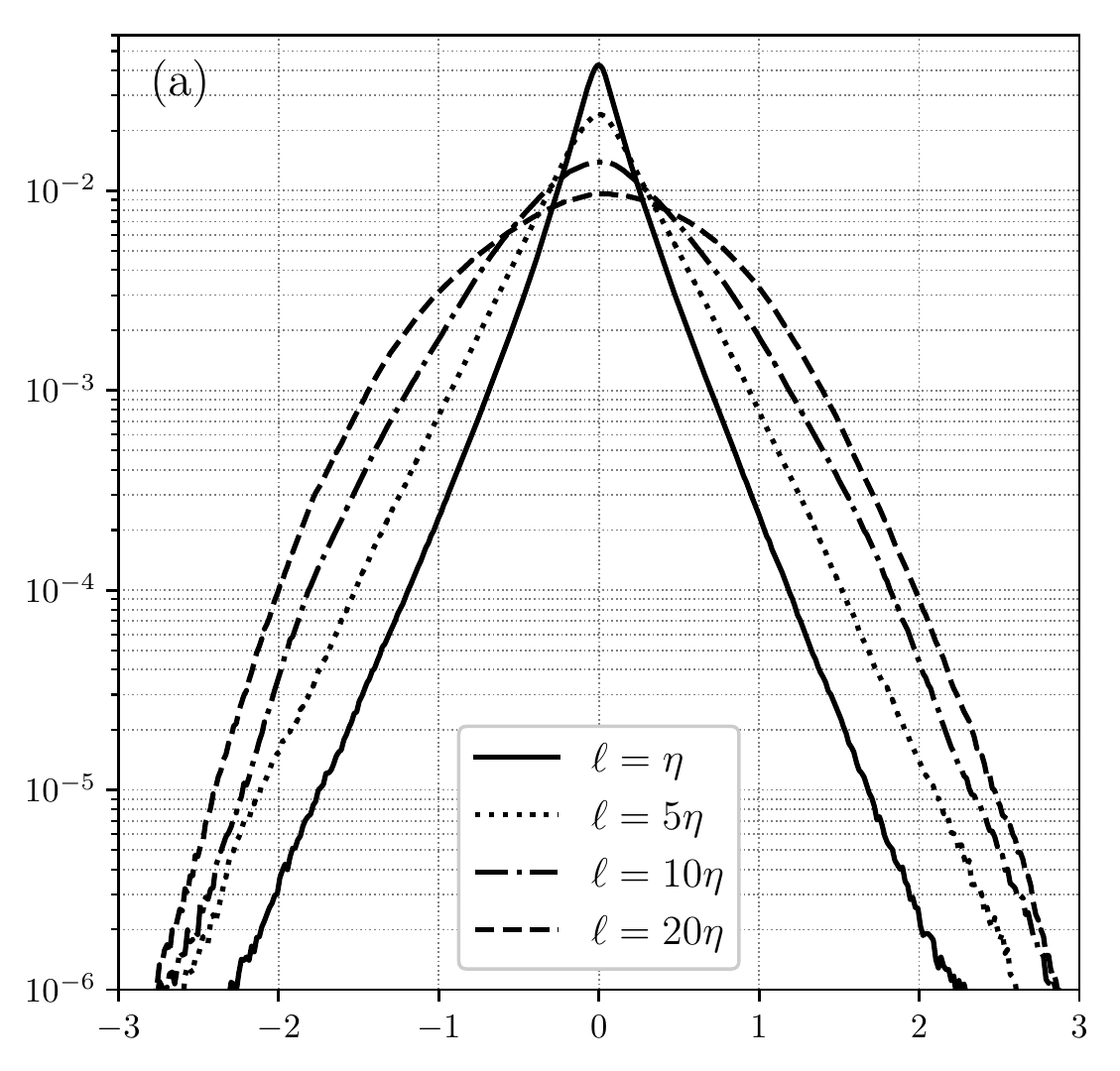} \\
	\includegraphics[width=.45\textwidth]{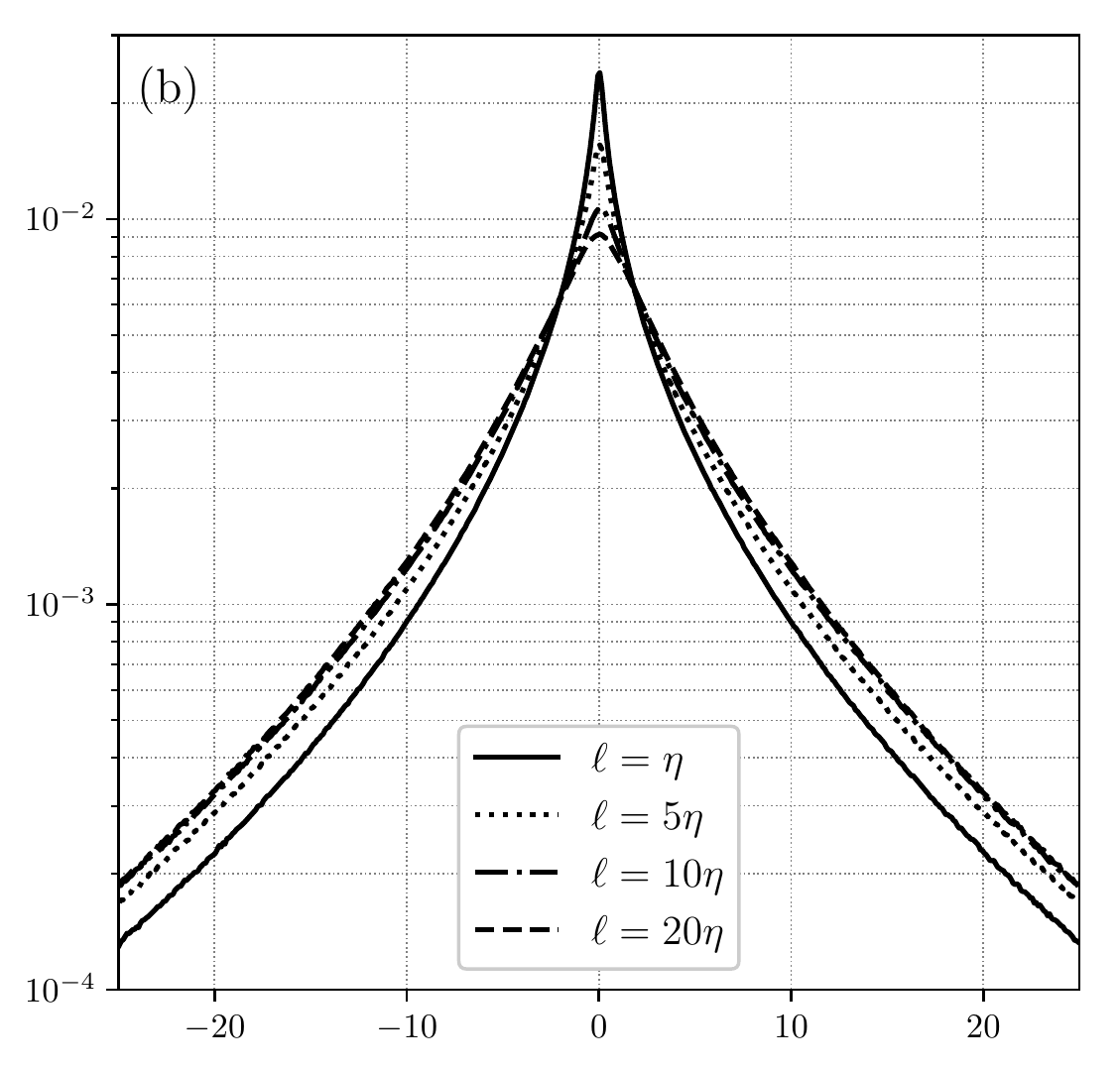}
\end{center}
\caption{For Run IV, PDFs of the transverse velocity {(a) and Lamb (b)} increments with $\ell=\eta$, $\ell=5\eta$, $\ell=10\eta$ and $\ell=20\eta$, where $\eta$ is the Kolmogorov dissipation scale.}\label{fig3}
\end{figure}

\begin{figure}
\begin{center}
	\includegraphics[width=.48\textwidth]{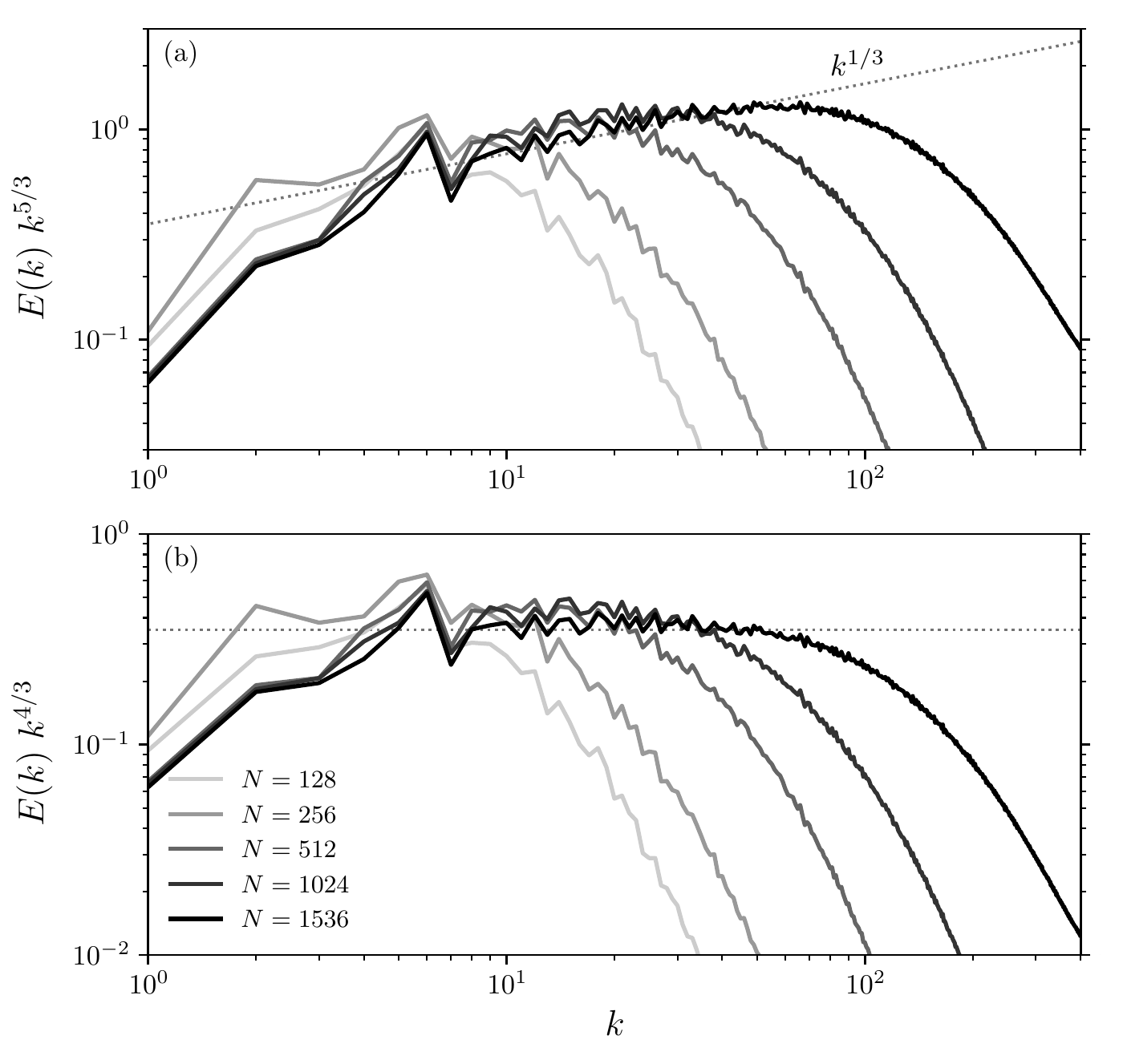}
\end{center}
\caption{Energy spectra for all Runs in Table \ref{table} as a function of wavenumber $k$.}\label{fig4}
\end{figure}

\begin{figure}
\begin{center}
	\includegraphics[width=.45\textwidth]{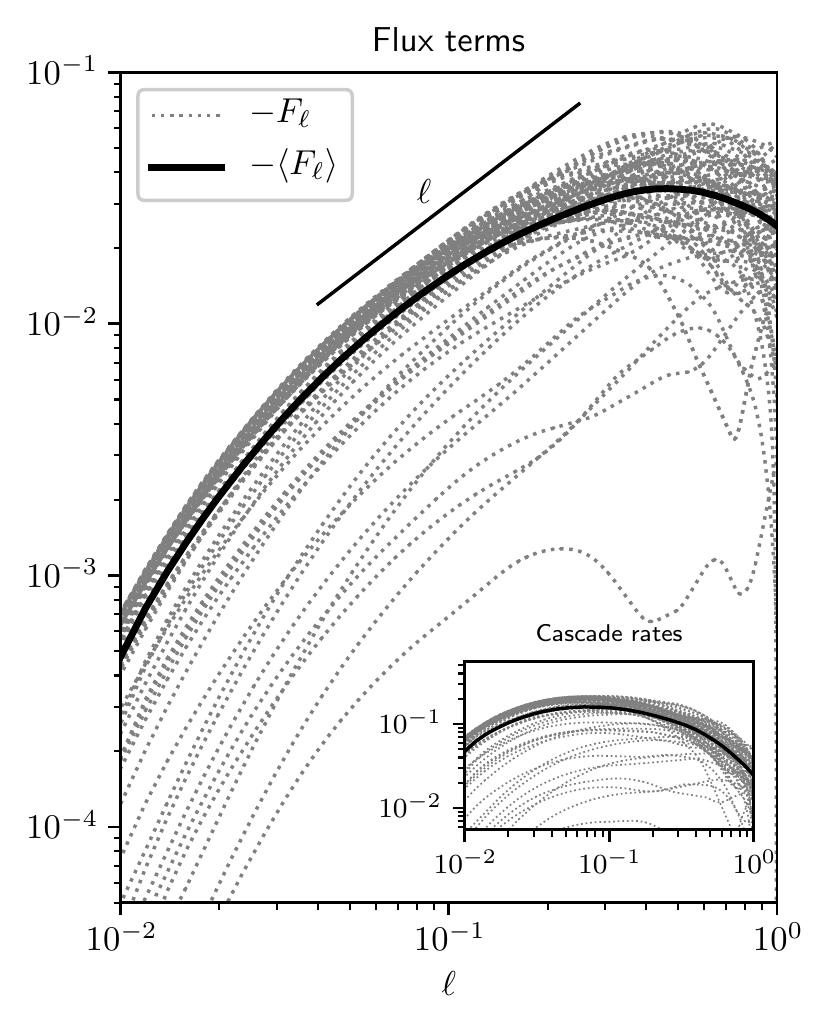}
\end{center}
\caption{For Run IV: mean structure function $\langle F_\ell \rangle$ (black-solid) and structure functions $F_\ell$ along different increment directions $\boldsymbol{\ell}$ (gray-dot) using Eq.~\eqref{projection}. Inset plot: energy  cascade rate along different directions (gray-dot) and mean cascade rate (black-solid).}\label{fig5}
\end{figure}

\section{Results}\label{results}

\subsection{Statistical dynamics of the velocity, vorticity and Lamb vector}\label{statistics}

The Lamb vector is known to be of great importance for fluid dynamics. In particular, it is essential for the nonlinear dynamics of turbulence \citep{C2015,T1990} since the nonlinear term in the Navier-Stokes equation \eqref{n-s} can be written as a function of the Lamb vector cross product the velocity vector plus a gradient term. Then, in order to study the turbulent regime, we discuss the statistics properties of the velocity, vorticity and Lamb vectors.

Fig.~\ref{fig1} shows three snapshot of the velocity (a), vorticity (b) and Lamb vector (b) modulus for Run V at the time when the dissipation reaches its maximum value. In the three panels, the large-scale structures are a signature of the initial condition (see Sec. \ref{equations}), while the small-scale structures are produced by the nonlinear dynamics and the direct cascade of energy. As we expect, since the Lamb vector is the cross product between the velocity and vorticity fields, it shows a chaotic, multi-scale and intermittent behaviour (in which strong gradients are highly localized). 
 
Several statistical features associated with isotropic and homogeneous turbulence can be observed from our numerical results. Fig.~\ref{fig2} shows the probability distribution functions (PDFs) for the velocity (a), vorticity (b) and Lamb vector (c) components, for Run IV. While each velocity field component shows a clear Gaussian distribution with an approximate zero mean value, the vorticity and Lamb vector components show a more exponential or peak distribution. The Lamb vector statistical behaviour is a direct consequence of the vorticity field dynamics in homogeneous turbulence \citep{T1990}. In particular, a more direct approach to characterize a turbulent flow, is to compare the PDFs of velocity increments at different two-point distances $\ell$. Then, we can defined the parallel and perpendicular velocity increment as,
\ba 
\delta {u}_\parallel &=& \boldsymbol\ell\cdot[{\bf u}({\bf x} + \boldsymbol\ell) - {\bf u}({\bf x})], \\
\delta {\bf u}_\perp &=& \boldsymbol\ell\times[{\bf u}({\bf x} + \boldsymbol\ell) - {\bf u}({\bf x})].
\ea
{Fig.~\ref{fig3} shows the PDFs for $u_\perp=|{\bf u}_\perp |$ {(a) and ${\cal L}_\perp= |\boldsymbol{\cal L}_\perp |$ (b) increments} for different separation distances $\ell$. For large separation distances, for the $\delta u_\perp$ we observe distributions close to the Gaussian distribution with decaying tails (i.e., presence of strong gradients). On the other hand, as we expect for a turbulent and intermittent fluid, Fig.~\ref{fig3} {(a)} shows the development of exponential and stretched exponential tails as the increment separation distance $\ell$ decreases. It is worth mentioning that this behaviour is not observed in the perpendicular Lamb vector increments. In particular, we observe exponential or peak distributions for all scale separations.}

{Fig. \ref{fig4} shows the kinetic energy spectra compensated by (a) $k^{5/3}$ and (b) $k^{4/3}$  as a function of the wavenumber $k$, for all Runs in Table \ref{table}. Typically, in incompressible HD turbulence, an inertial range corresponds to Kolmogorov-like $k^{-5/3}$ scaling. However, our numerical results show a scaling close to $k^{-4/3}$ instead. This behaviour has already been reported in \citet{Mi2006a}, where a difference of $1/3$ was found in the scaling of kinetic energy spectrum. This departure is most likely due to the bottle-neck effect \citep[see, e.g.][]{Mi2006a,K2004}. Nevertheless, the bottle-neck effect was found to be prominent mostly for 3D simulations with grid points below $1024^3$. In our present study, the -4/3 slope is still present in $1536^3$ grid points. It is worth noting that previous studies also reported a departure of the spectral index by 0.1 due to intermittency effects \citep{K2003}. In a more general sense, we interpreted our numerical results as a combined effect of bottle-neck and intermittency. A detailed discussion on this subject is, however, is beyond the scope of the present work where the velocity power spectra are drawn only to get a prior idea of the inertial zone in $k$-space ($\sim 8\times10^0 - 6\times10^1$ for Run V).}


\begin{figure*}[t]
\begin{center}
	\includegraphics[width=.8\textwidth]{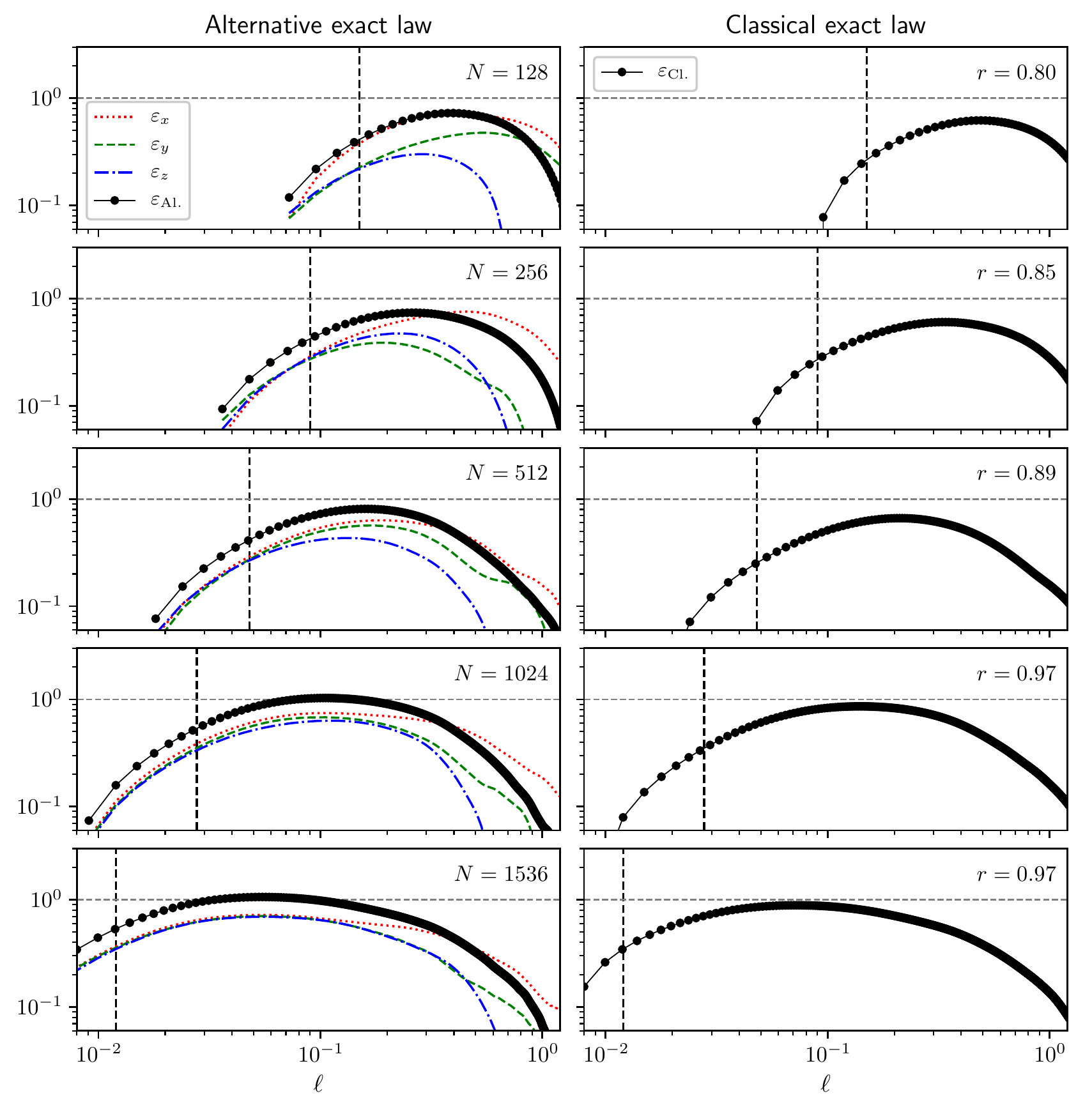}
\end{center}
\caption{Energy cascade rates using Eq.~\eqref{first1} (left panel) and using Eq.~\eqref{projection} (right panel) as a function of $\ell$, for all Runs.}\label{fig6}
\end{figure*}

\subsection{Computation of velocity and mixed structure functions}\label{corr}

For the computation of velocity and mixed structure functions in multiple directions (and thus to obtain statistical convergence by averaging over all these directions), we use the angle-averaged technique presented in \citet{Ta2003}. This technique avoids the need to use 3D interpolations to compute the correlation functions in directions for which the evaluation points do not lie on grid points. This  significantly reduces the computational cost of any geometrical decomposition of the flow \citep{Ma2010}. In particular, we have used a decomposition based in the SO(3) rotation group for isotropic turbulence \citep[see,][]{B2001,A2018}. 

The procedure used to compute each term in the exact law given in Eq.~\eqref{first1} (or Eq.~\eqref{projection}) over several directions can be summarized as follows: in the isotropic SO(3) decomposition, the mixed structure functions are computed along different directions generated by the vectors (all in units of grid points in the simulation box) (1,0,0), (1,1,0), (1,1,1), (2,1,0), (2,1,1), (2,2,1), (3,1,0), (3,1,1) and those generated by taking all the index and sign permutations of the three spatial coordinates (and removing any vector that is a positive or negative multiple of any other vector in the set) \citep[see,][]{A1999,Ta2003}. This procedure generates 73 unique directions. In this manner, the SO(3) decomposition gives the mixed structure functions as a function of $73$ radial directions covering the sphere \citep{Ta2003}. The average over all these directions results in the isotropic mixed structure functions which depend solely on $\ell$.

As an example, Fig. \ref{fig5} shows the third-order structure function $F_\ell = |\delta{\bf u}|^2\delta u_\parallel$ for the 73 different directions in gray-dot line (for Run IV). Overplot is the average structure function $\langle F_\ell \rangle$ in black-solid line. {Inset plot is the energy cascade rate $\varepsilon$,}
{
\begin{equation}
    \varepsilon = -\frac{3}{4}\frac{\langle|\delta{\bf u}|^2\delta u_\parallel\rangle}{\ell}.
\end{equation}i
}
{On the other hand, from the alternative exact law \eqref{first1}, $\varepsilon$ is simply the average second-order mixed correlation function between the velocity and Lamb vectors divided by 2. It is worth mentioning that the computation of $\varepsilon$ using in situ measurements and Eq.~\eqref{first1} (i.e., the computation of the vorticity field) can be achieved using multispacecraft techniques, as the curlometer technique \citep[e.g., see,][]{D2002}. In general, this technique requires simultaneous measurements from four spacecrafts to be able to compute gradients. In particular, this technique have been used to compute electric currents and vorticity fields with in situ observation from Cluster and the most recent NASA Magnetospheric Multiscale (MMS) mission.} In the next Sec. \ref{energy}, we use the technique describe above to compute the energy cascade rates for all Runs in Table \ref{table} according to the alternative \eqref{first1} and the classical \eqref{projection} exact laws.

\subsection{Energy cascade rates}\label{energy}

Fig. \ref{fig6} shows the energy cascade rates as a function of the two-point distance for each run in Table \ref{table} using the alternative and the well-known Kolmogorov-Monin form. In the left panel we plot {$\varepsilon_\text{Al.}$} using the alternative exact law \eqref{first1} (black-solid) and its components \eqref{epsx} (red-dot) \eqref{epsy} (green-dashed) and \eqref{epsz} (blue-dot-dashed) and in the right panel we plot the energy cascade rate {$\varepsilon_\text{Cl.}$} using Eq.~\eqref{projection}. In vertical black-dashed line is the Taylor scale. The integral scale for each Run is larger than 1.25, i.e. the upper x-axis limit. Each plot in Fig.~\ref{fig6} have been normalized to their corresponding energy dissipation rate $\nu\langle\omega^2\rangle$ (see Table \ref{table}). {Finally, for each Run, we report the mean ratio $r=\varepsilon_\text{Cl.}/\varepsilon_\text{Al.}$, where the average has been computed along each inertial range.}

When we increase the spatial resolution we obtain a flatter region where the total energy cascade rate is constant thereby corresponding to the inertial range. In particular, for the largest spatial resolutions, i.e. $N=1024$ and $N=1536$, the inertial range obtained from the classical exact law is quite similar to the one obtained from the alternative exact law {($r=0.97$ in both cases)}. Moreover, in contrast to Eq.~\eqref{original} where we had to project the local divergence operator in the direction of $\boldsymbol\ell$, using Eq.~\eqref{first1}, {$\varepsilon_\text{Al.}$} was obtained directly from the measurements of the scalar product of the Lamb vector increments with the velocity field increments. This is clearly an improvement with respect to the old formulation of the exact relations \citep{Ga2009} and, in addition, it would be very efficient to compute energy cascade rates in turbulent systems where there is a privileged direction (e.g., turbulence with rotation or with a background magnetic field).

\section{Discussion and Conclusions}\label{discussion}

To the best of our knowledge, this is the first time that the alternative exact law Eq.~\eqref{first1} is numerically validated. Using a SO(3) isotropic decomposition, we have computed the energy cascade rate and we have investigated the statistical properties of the velocity, vorticity and Lamb vector for freely decaying homogeneous turbulence. For different spatial resolutions, our numerical results show that the energy cascade rate can be obtained directly from the measurements of the scalar product of the Lamb vector increments $\delta{\bf\cal L}$ with the velocity field increments $\delta\uu$. This indeed provides an advantage over the tradition Kolmogorov-Monin differential form which need to be integrated to compute $\varepsilon$.

We have studied several features associated with isotropic and homogeneous turbulence. In particular, the PDFs for the velocity components show a clear Gaussian distribution with a zero mean value whereas both the vorticity and the Lamb vector components show exponential or peak distribution. Moreover, the PDFs for the velocity increments for large separation distances show distributions close to Gaussian, while we observe the development of exponential and stretched exponential tails as the increment distance $\ell$ decreases, a direct consequence of the presence of intermittency in the fluid.  

For the largest spatial resolutions, we observe similar inertial ranges obtained from the classical exact law or the new alternative exact law. As we discussed before, this is a clear advantage of the alternative exact law since to be able to use Eq.~\eqref{projection} is mandatory to project the local divergence operator into the increment direction $\boldsymbol\ell$, while the energy cascade rate obtained from Eq.~\eqref{first1} is obtained simply from the measurements of the scalar product of the Lamb vector increments with the velocity field increments. 

Finally, as we increase the spatial resolution, we observe that the three correlation function components in Eq.~\eqref{first1}, i.e. $\varepsilon_x/2$, $\varepsilon_y/2$ and $\varepsilon_z/2$, converge to one-third of the total energy cascade rate $\varepsilon$ in the inertial range. These results are a direct consequence of the isotropy in the system. As we reach the dissipation or the injection scales for each Run, the different contributions  $\varepsilon_x$, $\varepsilon_y$ and $\varepsilon_z$ separate from each other. {It is worth mentioning that in presence of anisotropy (strong magnetic field or a rotation axis) some ideal invariants of the system could be transferred to both large (the so-called inverse cascade) and small scales. In the case of rotating and/or stratified flows \citep{M2015,Su2016}, this new alternative methodology could be useful in the research of geophysical turbulent flows. An interesting question would be, how the three energy cascade components $\varepsilon_x, \varepsilon_y$ and $\varepsilon_z$ behave in a non-isotropic medium? In part, this question will be addressed elsewhere in which we include a strong magnetic guide field into the system.}

\section*{Acknowledgments}

N.A. is supported through a DIM-ACAV post-doctoral fellowship. N.A. acknowledge financial support from CNRS/CONICET-UBA Laboratoire International Associ\'e (LIA) MAGNETO. S.B. acknowledge support from DST INSPIRE research grant. The authors acknowledge Sebastien Galtier for useful discussions.

\bibliographystyle{apsrev4-1}
%

\end{document}